# Polaronic Contributions to Friction in a Manganite Thin Film

*Niklas A. Weber[1], Dr. Hendrik Schmidt[1], Tim Sievert[1], Prof. Christian Jooss[1], Dr. Friedrich Güthoff[2], Prof. Vasily Moshneaga[3], Prof. Konrad Samwer[3], Prof. Matthias Krüger[4], and Prof. Cynthia A. Volkert[1,5,\*]*

[1] *Institute of Materials Physics, University of Göttingen, 37077 Göttingen, Germany*

[2] *Institute of Physical Chemistry, University of Göttingen, 37077 Göttingen, Germany*

[3] *1. Physics Institute, University of Göttingen, 37077 Göttingen, Germany*

[4] *Institute for Theoretical Physics, University of Göttingen, 37077 Göttingen, Germany*

[5] *The International Center for Advanced Studies of Energy Conversion (ICASEC), University of Göttingen, 37077 Göttingen, Germany.*

**E-mail:** volkert@ump.gwdg.de

Despite the huge importance of friction in regulating movement in all natural and technological processes, the mechanisms underlying dissipation at a sliding contact are still a matter of debate. Attempts to explain the dependence of measured frictional losses at nanoscale contacts on the electronic degrees of freedom of the surrounding materials have so far been controversial. Here, it is proposed that friction can be explained by considering damping of stick-slip pulses in a sliding contact. Based on friction force microscopy studies of $La_{(1-x)}Sr_xMnO_3$ films at the ferromagnetic-metallic to paramagnetic-polaronic conductor phase transition, it is confirmed that the sliding contact generates thermally-activated slip pulses in the nanoscale contact, and argued that these are damped by direct coupling into phonon bath. Electron-phonon coupling leads to the formation of Jahn-Teller polarons and a clear increase in friction in the high temperature phase. There is no evidence for direct electronic drag on the atomic force microscope tip nor any indication of contributions from electrostatic forces. This intuitive scenario, that friction is governed by the damping of surface vibrational excitations, provides a basis for reconciling controversies in literature studies as well as suggesting possible tactics for controlling friction.

## 1. Introduction

Understanding and controlling friction is a long standing, major topic in both research and application. Earliest approaches to controlling friction have focused on changing the sliding contact by modifying surface roughness or adding lubricants. More recently, atomic force microscope (AFM) methods have allowed basic studies of friction which intensively address how energy is dissipated at a sliding contact.[1–4] In addition to the atomic interactions in the contact interface, it has been found that phonon dissipation mechanisms,[5] conduction electron excitation and drag,[6] and electrostatic forces[7–9] in the surrounding materials can significantly influence the friction between two bodies. This offers the intriguing possibility to change friction independent of the exact nature of the contact interface, which is strongly affected by operating conditions. Eventually, the goal is to bridge from the nanoscale contacts of AFM-based studies to macroscale contacts, which are composed of many interacting nanoscale contacts, with the hope of developing a physical basis for optimizing engineering friction.

Although our understanding of the basic mechanisms of sliding friction is surprisingly vague, there are two remarkable characteristics of friction which we now understand. The first is the century-old observation that the friction force $F_f$ on a macroscopic object is proportional to the normal force $F_N$ between the object and the surface on which it moves and not to the macroscopic contact area.[10] This at-first confusing observation seems to contradict the basics of classical mechanics, but was finally resolved by the understanding that the true contact area $A$ between two objects is composed of many nano-asperities, and is much smaller than the apparent contact area; the true contact also increases roughly linearly with the normal force.[10,11] Since then, several studies have indeed confirmed that the friction forces at both single asperity nano-contacts and multi-asperity macro-contacts scale with the true contact area,[12,16] so that

$$F_f = \tau A, \tag{1}$$

where $\tau$ is a shear stress which is required to generate and propagate the excitations that allow sliding in the contact area.[17] It is conceptually analogous to the Peierls stress required for dislocation glide[18,19] but is expected to be smaller due to slightly increased interatomic distances due to incommensurability and defects and the presence of (often amorphous) oxide and contamination layers.

The last several decades have shown great progress in using AFM to identify the excitations that are stimulated at nanoscale sliding contacts. It is widely observed in both experiment and theory that sliding contact does not proceed continuously, but by stick-slip event.[14,20-23] The stick-slip events are observed at all length scales, accounting for earthquakes[24] and atomic scale instabilities,[20] as well as the acoustics of violins, squeaky doors, grasshoppers, and squealing brakes. With the exception of contacts containing only a few molecules, interface sliding should be envisioned as the motion of slip pulses through the contact interface, where the local slip velocity far exceeds the average slip velocity of the contact.[19] Even cases of apparent continuous sliding are often attributed to local slip pulses moving through the interface.[25] Thus, the sliding velocity of an AFM tip for the one dimensional case can be expressed as

$$v = A\rho v_{sp} \tag{2}$$

where $A$ is the contact area, $\rho$ is the areal density of active slip pulses (so that $A\rho$ is the number of active slip pulses in the contact), and $v_{SP}$ is the slip pulse velocity. This equation is analogous to Orowan's equation for the strain rate due to dislocation glide. However, in contrast to the situation for crystal lattice dislocations, there are no crystal topology constraints on slip pulses in a general incommensurate interface. In particular, the expected local variations in atomic structure at a general two dimensional nanoscale contact allow the forward slip distance of a slip pulse (analogous to the Burgers vector) to vary from position to position within the contact. Furthermore, the slip pulse is primarily a dynamic defect in that it does not have a clear structural signature while at rest, although it presumably forms in regions where the local atomic arrangements at the interface make for easy slip. We note that not all three-dimensional Molecular Dynamics simulations of sliding friction show evidence of slip pulse generation, which may be because of the high speeds often used in such studies.[26]

Stick-slip behavior at the atomic scale,[27-29] through the mesoscale,[22] to the macroscale[23] has been successfully modeled using minimal one-dimensional models such as the Prandtl-Tomlinson and Frenkel-Kontorova models. The remarkable success of these models lies in the fact that they can account for the strongly nonlinear dependence of friction on sliding velocity, which is the second remarkable characteristic of friction. By describing the motion of an AFM tip as spring-loaded thermally-activated motion over an energy barrier, quantitative agreement has been obtained with sliding velocity dependences, provided the ratio of barrier height to spring stiffness is high enough that an instability occurs.[28-33] In essence, the models are based on an Arrhenius-type law for the one-dimensional slip pulse velocity of the form[30,31]

$$v_{sp} = \omega_o L exp\left(\frac{-\Delta E}{kT}\right) \quad (3)$$

where $\omega_0$ is an attempt frequency, $L$ is a factor with units of length that depends on the contact stiffness of the experimental set-up and temperature,[30,31] and $\Delta E$ is the activation energy barrier which must be overcome for slip to occur. The basic assumptions that lead to this equation are that a slip event is critically damped, meaning that neither reverse slip jumps (overdamped) nor multiple jumps from inertial effects (underdamped) occur. There is evidence from atomic stick-slip behavior to support this idea,[27] and near-critical behavior may in fact be an emergent property in a sliding contact.[34]

Assuming that the activation barrier for slip $\Delta E$ is reduced from the unloaded value $E_0$ of the activation barrier by the force $F_{SP}$ driving the slip event,[30,31] an equation for the friction force can be obtained that captures the basic features of the velocity dependence, namely that the friction force is finite in the limit of zero sliding velocity and shows a weak dependence on sliding velocity. A number of different approximations and validity ranges have been considered,[26,28-33] all leading to a logarithmic dependence of the friction force on the sliding velocity. Using a linear approximation $\Delta E = E_0 - bF_{SP}\pi^{-1}$ to describe the dependence of the activation barrier on the force acting on the slip pulse $F_{SP}$,[30] one obtains an equation that approximately captures the dependence of the friction force on sliding velocity,

$$F_f = F_{ph} = \frac{\pi A}{b^2 w}\left[E_o + kT ln\left(\frac{v}{v_o}\right)\right]. \quad (4)$$

We have used **Equation (1)** to **Equation (3)**, set $v_0 = A\rho\omega_0 L$, and related $F_{SP}$ to the shear stress $\tau$ in the contact interface as $F_f = \tau b w$ where $b$ is the mean forward slip distance of the slip pulse. This one-dimensional model describes a slip pulse that moves forward in a simultaneous jump across the entire width $w$ of the contact. Note that the energy barrier $E_0$ and shear stress $\tau$ will increase with the normal force,[31] analogous to the dependence of dislocation glide on normal force, so that the friction force $F_f$ depends on the normal force $F_N$ both through the contact area $A$ and $E_0$.

Equations for thermally-activated stick-slip behavior with forms similar to **Equation (4)** have been widely used in the literature to discuss the velocity dependence of friction.[28,30-33] The models are able to fit the data quite well, but surprisingly, predict that friction should only depend on surface structure (through $E_0$, $b$ and $\rho$) and contact area ($A$), and not on the inherent dissipation rates in the surrounding materials.[27] In reality, it is clear that vibration modes at surfaces couple directly to the macroscopic degrees of freedom of the underlying materials, as has been widely discussed within the context of adsorbate vibration relaxation.[35] The predicted independence from inherent dissipation within the Prandtl-Tomlinson model follows from the assumption of critical damping, which is supported by both experiment and simulation at the atomic scale.[27, 34] In fact, thorough modeling of friction for a wide range of dissipation rates show that the friction force increases with the dissipation rate in both the underdamped and overdamped regimes, but is relatively constant in the critically-damped regime,[29] in good agreement with Kramers reaction rate theory.[36] The question of which regime best describes experimental data remains open.

In fact, a number of experimental studies show a clear dependence of friction on the surrounding material properties,[37-43] suggesting that it is time to move beyond the widely used critically-damped Prandtl-Tomlinson model (e.g. Equation (3) and Equation (4)), despite its impressive success accounting for the velocity dependence of friction. The most compelling experimental studies have investigated friction across the superconducting transition. Despite relatively unchanged surface bonding and structure, a clear increase in friction on transitioning from the superconducting to normal state has been observed in a variety of materials.[40-43] One explanation given for the behavior is based on viscous damping of the tip by electronic excitations: in the superconducting state the electrons form Cooper pairs that exhibit an energy gap in their excitation spectrum, while in the normal state quasi-free electrons are easily excited and dissipate the sliding energy. An electronic friction force $F_{el}$ is simply added to the total friction force as $F_f = F_{ph} + F_{el}$, where the phononic friction $F_{ph}$ results from the slip pulse excitations described above in Equation (4). The electronic friction force $F_{el} = -B_{el}v$ is assumed to represent damping by normal carriers near the Fermi energy through a viscous damping coefficient $B_{el}$ as a result of electromagnetic interactions with the tip.[40,42,43] However, theoretical values for the damping coefficient based on generating electronic excitations near the Fermi energy and dragging them behind the tip[1] deliver numbers that are much too small to account for the measured changes in the sliding friction force.[40,42] On the other hand, there is good agreement with the much smaller forces measured in non-contact friction studies,[43] where presumably the slip pulse excitations described in Equation (4) are not active. This suggests that viscous electronic damping of van der Waals interactions with the tip is active and presumably contributes to sliding contact friction; it is simply completely overshadowed by the energy dissipated in the slip pulse excitations. This same problem with the order of magnitude of possible contributions from electronic excitations has been encountered when trying to explain the effect of doping and carrier density on the sliding friction of various semiconductors.[7,39]

Since the energies required to generate electronic excitations through electrodynamic interactions with the sliding tip or to drag image charges (Coulomb drag) are both too small to explain the observed correlation between material properties and contact sliding friction, the effect of electrostatic forces has also been considered.[7,37,39,40,42] Forces both parallel and perpendicular to the surface normal can be generated depending on the distribution of charges or dipoles in the contacting materials. Electrostatic normal forces have the effect of increasing $A$ in Equation (1) as a result of elastic contact mechanics (e.g. Hertz contact theory), while net electrostatic forces parallel to the sample surface will directly add to the friction forces. In some cases, an order-of-magnitude agreement with experiment has been achieved, but often quite large and specific trapped charge densities have been required. For example, in order for electrostatic forces to contribute to the lateral forces felt by the tip, the "line of charges" left in the wake of the sliding tip must have lifetimes long enough to slow the advancing tip but short enough to have annihilated by the time the tip is rescanned along the same path.[7,37] Such highly specific scenarios have not yet been verified, although they would offer promising methods to control friction by tailoring electric fields.[44]

In this paper, we investigate nanoscale single asperity sliding friction at the surface of La$_{(1-x)}$Sr$_x$MnO$_3$ (LSMO) films ($x$ = 0.2 and 0.3) while heating through transitions from the ferromagnetic metal (FM) to a paramagnetic polaronic conductor (PM) state. We use experimental conditions where the contact contains many atoms (>100), rather than attempting single atom contacts, in order to probe the behavior of typical nano-asperities that make up macroscopic contacts. We observe a clear increase in friction on crossing from the metallic to small polaron hopping conductivity state, which is due to damping of the slip pulse excitations generated in the sliding contact. We argue that the strong increase in electron-phonon coupling at the transition causes a change in the lifetime of the slip pulses and can quantitatively account for the excess friction in the polaronic phase. This explanation is distinct from many previous explanations, which were based on van der Waals and Ohmic losses and were unable to quantitatively account for the observed friction.

## 2 Results and Discussion

**2.1 Friction and adhesion forces at the phase transition.**

$La_{(1-x)}Sr_xMnO_3$ films with $x = 0.2$ and $x = 0.3$ were chosen for the friction studies because they manifest closely spaced temperature-driven transitions in electrical ($T_{MM}$) and magnetic ($T_C$) properties without changes in the bulk film crystal structure and in atomic bonding. At low temperatures, the films are ferromagnetic with metallic conduction governed by the double exchange (DE) mechanism. They become paramagnetic above the transition temperatures with strongly increased electrical resistivity due to the transition to a small polaron conductor. The strong increase in electron-phonon coupling results in localization of charge carriers at the Jahn-Teller distortions to form small polarons.[45-49] We note that the transition temperatures are dependent on the exact composition (including possible oxygen vacancies) and on stresses that result from the epitaxial relation with the substrate[50,51] and may be different at the surface of the films due to an observed reconstruction.[52] Often the two transitions overlap due to the coupling between spin, electron, and phonon degrees of freedom, but can also be shifted from each.[45-49]

A 6 nm thick film of $La_{0.7}Sr_{0.3}MnO_3$ ($x = 0.3$ specimen) was fabricated on a buffered $SrTiO_3$ substrate by metal-organic aerosol deposition,[53] with a ferromagnetic-paramagnetic transition at $T_C = 338\ K$ and a metal-like to hopping small polaron conductivity transition at $T_{MM} = 330$ K. A 70 nm thick film of $La_{0.8}Sr_{0.2}MnO_3$ ($x = 0.2$ specimen) was fabricated on $SrTiO_3$ by sputter deposition and exhibits a ferromagnetic-paramagnetic transition at $T_C = 220\ K$ and a metal-poloranic conductor transition at $T_{MM}$ = 187 K. The crystal structure is rhombohedral both above and below the transitions ($R\bar{3}c$ space group) and strained due to the epitaxial relation with the underlying substrates.[50] Measurements to determine crystal structure, film thickness, magnetic properties and resistivity are summarized in the Supporting Information for both films.

Nanoscale friction force measurements[54] were performed on heating through the transition temperature using AFM-based lateral force microscopy for a range of normal loads and scan speeds. The AFM studies were performed in a UHV environment in order to avoid complications from water condensation and other surface contamination. Details of the force calibration and scaling are described in the methods section. The normal forces $F_N$ between the conducting nanoscale Si tip and the film were kept below 30 nN to avoid detectable wear contributions. A typical friction loop for an applied normal force of 0.7 nN is shown in **Figure 1(a)**. Including the adhesion force (**Figure 1(e)**) gives an actual normal force of about 12 nN and a true contact area of $A$ = 5 nm², according to elastic Hertz contact theory and using the nominal tip radius of 10 nm (see Experimental Methods section and Supporting Information). According to Equation (1), the friction force then corresponds to an interface shear stress on the order of $\tau$ = 2 GPa, which is as expected, a factor of 5 smaller than the Peierls barrier in perovskite oxides.[55] The ~10% variations in the magnitude of the friction force during the forward and backward traces do not correlate with topography and are attributed to stick-slip events in the contact. To first approximation in the one-dimensional model used above, the forward slip distance of the stick-slip events is $b = 10\% \cdot \sqrt{4A/\pi}$ = 0.25 nm,[56] which is a reasonable number in that it is comparable to an interatomic distance.

Friction forces $F_f$ were obtained for each normal load and temperature by averaging the magnitude of the forces for many friction loops (**Figure 1(b),(c),(d)**). The friction forces for both films show a gradual decrease with increasing temperature followed by an abrupt increase at the transition from the metallic to polaronic conducting state. Note that the clear difference in the electrical and magnetic transition temperatures in the *x* = 0.2 film, allows us to identify that the friction change is correlated with the onset of polaron formation and not with a change in magnetic order. Depending on the normal force, the friction force then either falls again for low normal forces or stays at an elevated value for larger normal forces. Friction measurements recorded first and last in the temperature series were performed at the same temperature (room temperature for *x* = 0.3 and 250 K for *x* = 0.2) and show no significant difference relative to the error bars, so we assume no or only insignificant changes in the tip geometry or surface chemistry during the friction measurements.[57] Although the friction forces for the *x* = 0.2 film are smaller than for the *x* = 0.3 film (even for the same normal force), the fact that both films, with different compositions, different thicknesses, and different transition temperatures, both show the same behavior indicates that the correlation between the friction increase and the formation of small polarons is robust.

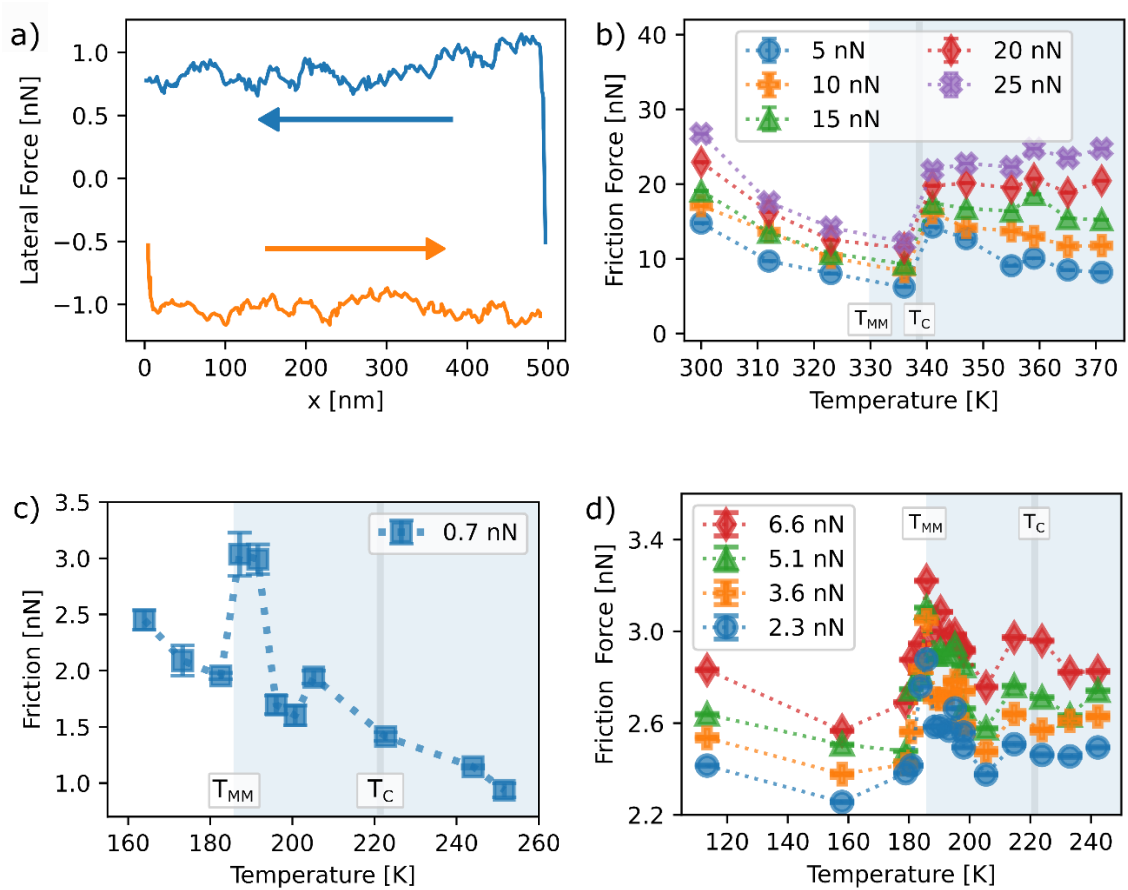

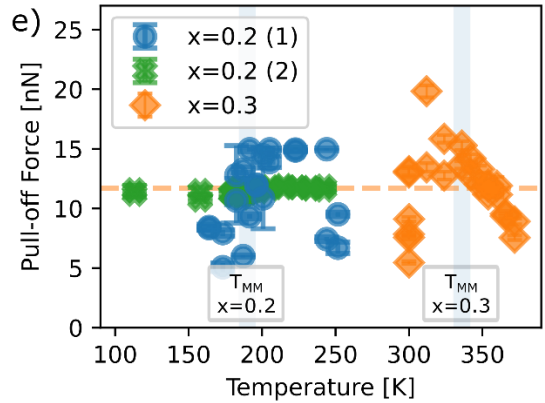

**Figure 1.** Friction and adhesion forces near the ferromagnetic metal to paramagnetic polaronic conductor transition for $x = 0.2$ and $x = 0.3$ LSMO films. (a) Lateral friction force loop for the $x = 0.2$ film obtained at a normal force of 0.7 nN and a sliding velocity of 2.5 µm s$^{-1}$ showing a strong hysteresis and stick-slip activity. (b) Average friction forces for the $x = 0.3$ film as a function of temperature for different normal forces at a sliding velocity of 2.5 µm s$^{-1}$. (c,d) Two measurements of average friction forces for the $x = 0.2$ film as a function of temperature for different normal forces at a sliding velocity of 0.25 µm s$^{-1}$. The measurement with a normal force of 0.7 nN (c) was performed separately using a different cantilever. (e) Adhesion forces estimated from pull-off force measurements for the $x = 0.3$ and $x = 0.2$ films.

The adhesive forces between the tip and LSMO films can be estimated by the pull-off forces that were obtained from force-distance measurements as a function of temperature. They are plotted in Figure 1(e) for both films and show no systematic trends with respect to the transition temperatures or the order of measurement. In particular, values recorded at the same temperature and first and last in the temperature series give the same value of the adhesive force, confirming that there was no measurable change in the tip geometry. The adhesion energy can be estimated as 90 mJm$^{-2}$ using the Derjaguin, Muller, and Toporov contact model,[58] which is consistent with typical interface energies measured in AFM studies.[59] The lack of an abrupt change in adhesion at the transition temperature rules out the possibility that the variations in friction are caused by changes in contact area. This is consistent with the fact that the elastic modulus changes by less than 5% at the transition temperature[60] and should lead to changes in the adhesion force on the order of 1 nN or less. Furthermore, the constant adhesion force confirms that neither the interatomic forces nor the electrostatic forces due to Coulomb and capacitive interactions change significantly at the phase transition. Thus, the observed changes in friction at the transition are due to a fundamental change in the ability of the near-surface regions of the films to dissipate energy.

**2.2 Friction coefficient and normal force dependence.**

The friction force is observed to increase linearly with the normal force at each temperature, with a non-zero intercept (**Figure 2(a)**). Even in the absence of an applied normal force ($F_N = 0$), an attractive force between the tip and specimen leads to friction forces between 5 and 15 nN and to the consequence that friction forces may be larger than the normal forces (see also Figure 1(a)). The observed linear dependence of the friction force on the normal force is in agreement with a number of other nanoscale friction studies,[4,5,9,61,62] although contact mechanical considerations for a sphere pressed against a flat surface predict sub-linear behavior, which is also often seen.[59,63] The widely observed linear behavior occurs for small adhesion energies[63,64] and may result from deviations from a spherical tip shape, from possible dependences of $\tau$ on the contact area size,[19] or simply from the combined effect of the normal force dependence of the contact area and the slip barrier (Equation (4)). The approximate linear dependence allows the microscopic definition of the friction coefficient (slope of the $F_f$ vs $F_N$ curve) to be used to characterize dissipation. The plot of the friction coefficient as a function of temperature (**Figure 2(b)**) shows two distinct trends. The friction coefficient gradually decreases with increasing temperature below the transition temperature and then abruptly increases at and above the transition temperature eventually leading to more than a doubling in value.

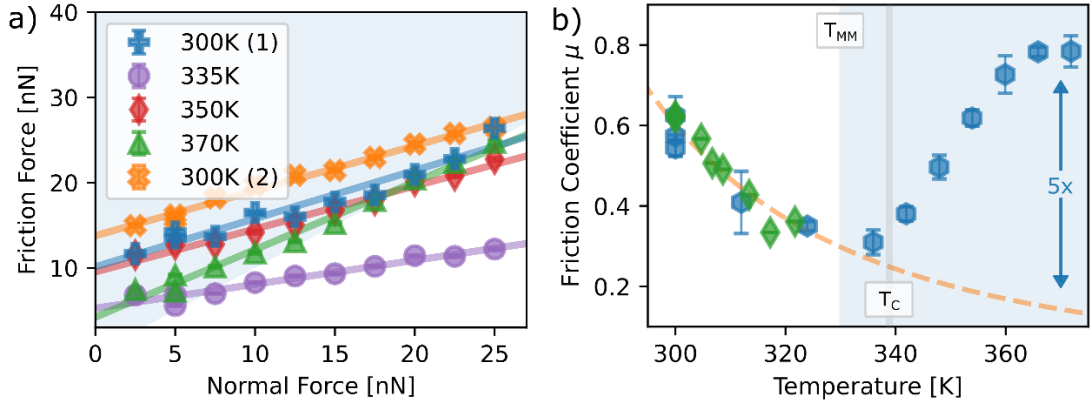

**Figure 2.** Friction of the *x*= 0.3 LSMO film. (a) Friction forces show a linear dependence on the normal force for all temperatures both above and below the transition. (b) The friction coefficient obtained from two sets of measurements with different cantilevers (blue diamonds and hexagons) from the slopes in (a) shows distinct behavior in the ferromagnetic metal and paramagnetic polaronic conductor states. The resistivity (green circles) in the polaronic phase is also shown. The dashed line shows a fit to the data with an Arrhenius dependence.

The gradual decrease in friction coefficient with increasing temperature is attributed to thermally activated stick-slip behavior.[28,29,32,42,63,66] This decrease is often successfully fit for small friction forces with an Arrhenius dependence $F_f = F_0 \exp(E_a/kT)$, where $E_a$ is assumed to be an activation barrier to slip.[27,57,65,66] The best fit to the data in the ferromagnetic state gives an activation energy of $E_a$ = 0.159 ± 0.013 eV , which is similar to the activation energy values obtained in previous studies on a range of different materials.[37,57] Note that Equation (4) predicts a linear or power law decrease of the friction force with temperature ($v < v_0$), which has also been widely applied in the literature.[31,33,42,66]

### 2.3 Velocity dependence of friction.

The friction forces were measured for both films as a function of tip sliding velocities (**Figure 3**) and show a clear non-linear increase with velocity, both above and below the transition temperature (**Figure 3(b)**). The velocity dependence is well described by a logarithmic dependence, as expected from thermally activated stick-slip models.[30-33,66] Fits were made to the data using Equation (4) and setting $E_0$ = 0.159 eV,[37] as obtained from the temperature dependence for the x=0.3 film. Otherwise the fits are under-constrained and do not stably converge. We obtain values between 1 nm s$^{-1}$ and 3 nm s$^{-1}$ for $v_o$, in good agreement with the literature,[63] and values for $b$ which increase with the temperature from 0.04 nm at 112 K to 0.14 nm at 300 K. These values are comparable to the values estimated from Figure 1(a). The idea that the slip distance might increase with temperature seems reasonable considering that the slip interface will include regions with low slip barriers which may be overcome by inertial processes.

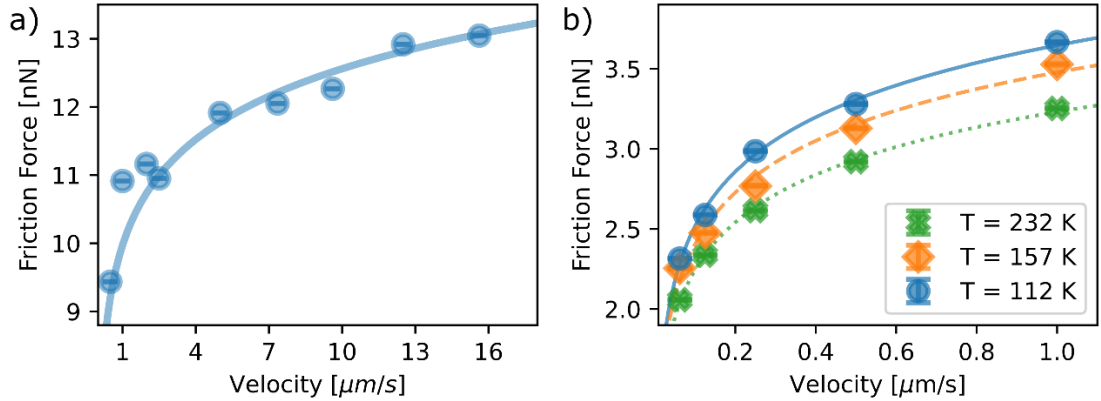

**Figure 3.** Friction forces as a function of scan velocity for the LSMO films. (a) Friction force in the $x$ = 0.3 film for $F_N$ = 5 nN at room temperature in the ferromagnetic state. (b) Friction force in the $x$ = 0.2 film for $F_N$ = 2.3 nN below, near, and above the transition temperature.

**2.4 Effect of electrostatic forces.**

Previous literature studies have proposed that electrostatic interactions between charges that are trapped near the specimen and tip surfaces might account for the observed dependence of friction on electronic properties.[7,37,39,40,42] For instance, the increase in friction at the insulator to metal transition of $VO_2$,[37,38] at the superconducting to normal transition of an oxide,[40,42] or as a result of changes in semiconductor doping,[7,39] have been discussed in terms of electrostatic forces resulting from charges or fields set up by the tip-specimen interactions. In some cases, the friction change has been accompanied by an adhesion change, lending quantitative support to the argument that electrostatic forces have increased the normal force and thereby increased the contact area.[7,37] In other cases, the required trapped charge densities are estimated, leading to reasonable numbers in the case of Si,[39] but requiring specific time-dependent distributions of charges in other cases.[7,37]

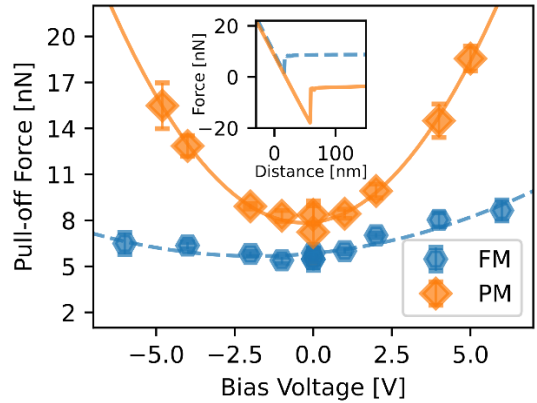

**Figure 4.** Pull-off forces for the *x* = 0.3 film as a function of bias voltage in the ferromagnetic metal (300 K) and paramagnetic polaronic conductor (380 K) states obtained from pull-off force measurements (see inset). The adhesion has a quadratic dependence on bias in both states (solid and dashed lines). The capacitance in the PM state is roughly 5 times larger than in the FM state. The contact potential difference between the AFM tip and specimen (given by the voltage at minimum adhesion) differs by approximately 2V between the two states.

In our studies, the lack of a temperature dependent adhesion force indicates that there are no significant differences between the Coulomb and capacitive interactions of the tip with the high and low temperature manganite phases under the conditions used to measure pull-off forces. This means that contact electrification (the charges transferred between two materials when brought together and then taken apart) is not measurably different for the two phases. On the other hand, it is well-known that the charges generated at two contacting surfaces depend on their sliding speed (tribocharging),[67] so that the possibility must be explored whether tribocharging is very different in the two phases. This possibility can be indirectly ruled out by considering the tribo-current that would be necessary above the transition to account for the measured change in friction. In Figure 1(b), we observe an increase in friction force of almost 10 nN at the transition. According to Figure 2(a), this requires an increase in normal force of about 20 nN. According to measurements of the pull-off force as a function of the applied bias voltage between the tip and the *x* = 0.3 film (**Figure 4**), a bias voltage in excess of 6 V would be needed to produce electrostatic forces of 20 nN between the tip and the PM phase. Such a large voltage is sufficient to cause resistive switching of LSMO.[68] Furthermore, conducting AFM measurements of the $x$ = 0.3 manganite film in both low and high temperature phases (see Supporting Information) show that currents of 100 nA would be generated by a bias voltage of 6 V. This would lead to a large current density of 2 MAcm$^{-2}$ at the contact, easily sufficient to produce large temperature rises. The behavior shown in Figure 1 allows us to rule this out. We thus find no support for the idea that electrostatic forces due to contact potential differences, contact electrification, or tribocharging can explain the observed increase in friction at the phase transition.

**2.5 Excess friction in the small polaron hopping conductivity state.**

The clear increase in friction at the transition from the metallic to polaronic state (Figure 1 and Figure 2) in both the $x$ = 0.2 and $x$ = 0.3 films shows that the excess friction results from the properties of the polaronic state. The absence of a corresponding increase in adhesion at the transition reveals that contributions from changes in the interatomic bonding or in the contact at the transition are insignificant, so that the excess friction must result directly from changes in the inherent damping of the near surface region of the LSMO. Further, the thermally activated behavior and the logarithmic dependence of the friction coefficient point to control by stick-slip events in both the metallic and polaronic conducting states (Figure 3). Thus, the interaction of the tip with the sample generates the same stick-slip pulse excitations above and below the transition, but these slip pulses are damped more strongly in the PM phase due to their interactions with small polarons. The idea of adding a viscous dissipation channel due to interactions of the tip with electronic degrees of freedom, as has been considered in previous publications (e.g. $F_f = F_{ph} + F_{el} = (\mu_{ph} + \mu_{el})F_N$[7,37,39,40,42]) is not relevant here. Furthermore, as has been noted previously, estimates of the magnitude of possible electronic contributions $F_{el}$, whether due to van der Waals interactions or to Ohmic losses, are several orders of magnitude too small to explain the observed changes in sliding friction both in the literature[7,37,39,40,42] and in our study.

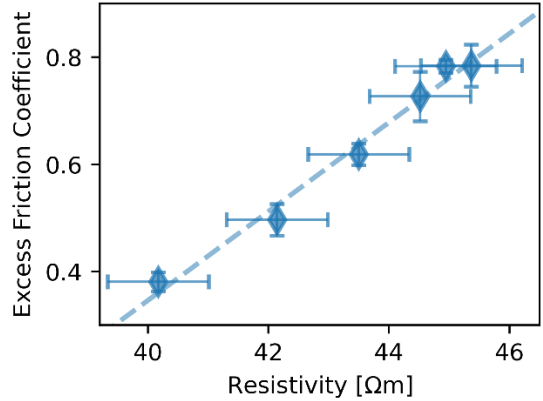

**Figure 5.** The slip pulse damping factor in the $x = 0.3$ film polaronic state scales with the resistivity.

We suggest instead that the friction force can be expressed as $F_f = \alpha_P F_M(T)$ where $F_M(T)$ is the friction force needed to generate slip pulse excitations in the metallic state (e.g. the dashed line in Figure 2(b)) and $\alpha_P$ is a slip pulse-polaron damping factor that reflects the strength of the slip pulse coupling to small polarons. The polarons that emerge around the transition temperature in LSMO consist of $e_g$ electrons localized on dynamic Jahn-Teller distorted $Mn^{3+}$.[45-49] This transition is also reflected in the electrical resistivity (Figure 1(b) and (d)), which is governed by the double-exchange mechanism well below the transition, but increases on passing through the transition as the Jahn-Teller energy increases and electron-phonon coupling becomes stronger.[45-49] The resistivity increases further with temperature until the electron-phonon coupling is strong enough that hopping polaronic conduction sets in with an accompanying decrease in resistivity (see Supporting Information).[46-49] In fact, the slip pulse-polaron damping factor defined above for the $x = 0.3$ film scales very well with the electrical resistivity in the neighborhood of the transition temperature where small polarons begin to form (Figure 5). This provides support for the idea that the vibrational slip-pulse excitations couple strongly to the phonon degrees of freedom, and through them, to the electron degrees of freedom as a result of the strong electron-phonon coupling which emerges around the transition. Coupling to the magnetic degrees of freedom will also occur but is expected to be weaker than the electron-phonon coupling effects in the temperature range investigated here[69]; furthermore, there is no evidence of an effect of the ferromagnetic transition in our studies (Figure 2(b)).

We are now faced with the question of whether the coupling of slip-pulses to polarons can account for an increase by a factor of 5 in the slip pulse damping $\alpha_P$ around the transition temperature (Figure 2(b)). The dominant phononic contribution to damping at the low velocities use for studying friction comes from the generation of sound waves,[70-74] as considered for the dissipation of adsorbate vibrations.[70,75] It is argued that the rate of frictional energy dissipation is proportional to $\omega^2$ or $\omega^4$ depending on whether the phonons with mode frequency $\omega$ are laterally coupled or oscillate independently, respectively.[76] In the case of $x$ = 0.3 material, a large change in the phonon spectrum is documented at the transition,[77,78] with the appearance of high wavenumber peaks associated with Jahn-Teller bond-stretching[77] appearing above the transition temperature. Assuming a complete shift of the populated phonon states from Mn-O bond excitations at $\hbar\omega \approx$ 47 meV[79] below the transition temperature to the Jahn-Teller peaks at $\hbar\omega \approx$ 62 meV and $\hbar\omega \approx$ 78 meV above the transition,[74,79] would result in an increase in friction by factors between 1.8 and 7.5, which could account for the observed friction increase by a factor of almost 5. It remains to argue that the slip pulses couple strongly to the Jahn-Teller modes. This will depend on complex details of the slip pulse excitations and the interatomic potentials. One possible reason for a strong coupling will be discussed in the following section. Another is based on the fact that slip pulses at the surface of LSMO are not only vibrational excitations. Due to the different interatomic bonding and valence states in LSMO, slip pulse motion will induce polarization and lead to dielectric excitations of the Mn-O octahedra and thereby to the Jahn-Teller modes.

**2.6 Slip pulse model for friction with explicit damping dependence.**

Given the success of the Prandtl-Tomlinson model in accounting for essential features of friction behavior, it is appealing to consider whether it can be extended to explicitly include damping of the slip pulse excitations. As already mentioned, most studies have focused on the critically-damped regime of spring-driven Prandtl-Tomlinson models where there is no explicit dependence of friction on damping.[27] Critically-damped behavior is supported by atomic stick-slip experiments which rarely show evidence of barrier re-crossing (overdamping) or inertial multi-slip (underdamping).[27] Furthermore, a recent simulation study[34] argues that phonon "dephasing"[75] in reaction to an atomic stick-slip event will lead to emergent critically-damped behavior. Which damping regime is relevant depends on the ratio of the inherent dissipation rate in the material $\eta$ to the local undamped contact resonance frequency $\sqrt{k/m}$, where $k$ is the contact stiffness and $m$ is an effective mass of the asperity. Critical damping occurs when the two rates are approximately equal, while underdamping and overdamping occur for smaller and larger inherent dissipation rates, respectively. A careful numerical study of the effect of damping on thermally activated stick-slip behavior shows that the friction force increases steeply with the dissipation rate in both the underdamped and overdamped regimes, and is relatively constant in the critically-damped regime.[29] Since local contact stiffnesses and effective masses depend on local atomic structure and normal stresses, all three damping behaviors may be simultaneously manifested in a general, incommensurate, nanoscale contact and an explicit dependence on the dissipation rate may be expected.

In this paper, we take a heuristic approach using guidance from the theory of dislocation glide to develop a model for nanoscale friction which explicitly includes inherent material dissipation.[19,72] In comparison with previously discussed scenarios, we take advantage of the two dimensional nature of the contact interface and the fact that the dynamic atomic-level processes for slip pulse motion may be different parallel and perpendicular to the slip direction, just as the motion of a dislocation is governed by overcoming the Peierls barrier to form a kink pair and then by the lateral motion of the kinks. The equation we obtain provides a framework to explain our observations, but we have not attempted a fully quantitative approach and do not claim the model is unique in being able to motivate our findings. However, the equation is successful in explaining the observed dependence of friction on the phase transition as well as on the temperature and velocity. Effectively, our equation describes underdamped Prandtl-Tomlinson behavior where the size of the slip step is controlled be viscously damped kink motion.

The well-studied example of dislocation glide in crystals provides a helpful analogy for understanding the atomic mechanisms and role of damping in the motion of a slip pulse. The essential feature is that dislocation glide is controlled by two distinct mechanisms: initiation of glide by kink pair nucleation, followed by lateral motion of the kinks (perpendicular to the slip direction) apart from each other. Even if it is clear that the topological constraints on dislocation structure, glide and kink motion do not apply to slip pulses in an incommensurate interface between two dissimilar materials, it is also clear that the two-dimensional nature of the contact interface requires mechanisms for propagation both parallel and perpendicular to the slip direction. For a disordered interface with large local variations in slip barriers, one might expect that slip pulses may disappear after propagating only a short distance along and perpendicular to the slip direction. In the limit that these distances are atomic distances, the resultant atomic stick-slip behavior provides the upper limit for the friction force at a fixed sliding velocity.[19] However, atomic stick slip is unlikely to be a major contributor to a general frictional contact, so that dislocation glide models based on non-adiabatic creation of kink pairs followed by viscously damped lateral motion of the kinks, may capture the essentials of slip pulse motion at a sliding interface.

The unloaded activation energy barriers for slip at an incommensurate interface between two dissimilar materials that are not chemically bonded ($E_o$ in Equation 4) are presumably quite a bit smaller than the Peierls barriers in single crystalline materials. Thus, given the large shear stresses under a sliding AFM tip (here, $\tau \approx F_f A^{-1} \approx$ 600 MPa), it is reasonable to assume that only forward moving kink pairs are generated in the sliding contact and that the large shear stress relative to the barrier height entirely hinders backward moving kink pairs. This is the same condition described by the critically-damped one-dimensional Prandtl-Tomlinson model for friction. Once a forward moving kink pair is nucleated, the large shear stresses in the interface will sweep the kinks laterally apart. Just as for dislocations, the barrier for lateral kink motion in a slip pulse is smaller than for kink nucleation. Therefore, we assume that kink velocity is controlled by drag, $v_k = \tau\, b\, w_k B_k^{-1}$, where $v_k$ is the kink pair separation velocity, $\tau$ is the shear stress acting on the kinks (assumed for simplicity to be the same as that driving kink pair nucleation), $w_k$ is the lateral distance swept out by the kink pair, and $B_k = \eta_k m$ is a kink pair viscous damping coefficient that is dependent on $T$.[80,81] The kink pair dissipation rate $\eta_k$ is the inverse of a kink pair lifetime. This then allows us to express the tip sliding velocity as the product of the rate of kink pair generation in the contact $A\,\rho\,\exp(-\Delta E/kT)$, the kink pair velocity $v_k$, and a factor $\alpha$ describing the contribution of each kink pair to forward sliding,[80,81]

$$v = \alpha \cdot v_k \cdot A\rho\, exp\left(\frac{-E_o + \frac{\Omega_{kp}}{\pi A}F_f}{kT}\right) = \frac{\beta F_f}{B_k} exp\left(\frac{-E_o + \frac{\Omega_{kp}}{\pi A}F_f}{kT}\right), \qquad (5)$$

where we have used equation (1) to equation (3) and replaced $b^2 w$ in the one-dimensional expression for $\Delta E$ (Equation 4) with the volume $\Omega_{kp}$ of the kink pair nucleus. Furthermore we have defined $β = αbw_{k}ρ$, which is a geometrical measure of the contribution of all kink pairs to forward sliding.[81] This equation has a similar form to Equation (3), but with the essential difference that the velocity pre-factor has now been replaced with a term proportional to the friction force and inversely proportional to a damping coefficient in the material. Just as for the Prandtl-Tomlinson model, this equation describes the generation of stick-slip events by non-adiabatic thermal activation of kink pairs, but in this case the contribution of each slip pulse to the sliding velocity is determined by the viscously damped motion of the kink pairs in the interface.

*Equation (5)* gives fits to the velocity dependence of the friction force (Figure 3) that are indistinguishable from those of Equation (4), but the interpretation of the fitting parameters is different. In contrast to the temperature dependence of $b$ obtained from fitting with Equation (4), the value of $Ω_{kp}$ from the fit with Equation (5) is roughly temperature independent, which seems physically reasonable. Meanwhile, $b$ increases by a factor of around 4 as the temperature increases from 112 to 232 K, consistent with the idea that the damping increases in the polaronic phase by the same factor as the friction. Except at low temperatures, the dominant dislocation drag effects are due to the phonon subsystem which generally behaves viscously.[72,82] The largest phonon loss contribution at low dislocation glide velocities is due to the generation of sound waves (phonon radiation friction), and has been estimated for the case of kinks moving along straight dislocations.[82] According to phonon coupling, a kink of width $D$ will couple most strongly into phonons with wavelengths $\gtrsim 2D$.[82] Since the barrier to kink motion is assumed to be small, $D$ will be larger than the atomic spacing, so that the main coupling to the Jahn-Teller modes of the Mn-O octahedral occurs via electron-phonon coupling.

We note that in addition to the explicit appearance of a viscous damping coefficient in Equation (5), there is also the a dependence on the friction force $F_f$ in the pre-factor. This gives the velocity equation features of both strongly driven systems (seen as a reduction of activation barrier by the friction force) and a weakly driven system (linear dependence on the driving force). Both the linear and exponential dependence of the tip velocity $v$ on the friction force $F_f$ result in a range of possible behaviors. At constant temperature, the velocity is more sensitive to changes in the friction force through the exponential term than through the linear term, explaining the observed logarithmic dependence of the friction force on velocity (Figure 3).

## 3 Conclusion

We report a clear increase in friction of two different LSMO films that correlates with the formation of Jahn-Teller distortions and small polarons. We are able to reproduce the temperature and velocity dependence of the friction and provide an order of magnitude justification based on the picture that the slip pulses generated in the sliding interface are damped by coupling to the phonon bath. In the case of the LSMO, this coupling increases at the transition temperature due to the emergence of small polarons in the near surface region of the material. We believe this model, which combines the widely accepted stick-slip behavior at sliding contacts with concepts for damping of moving dislocations, provides a solid basis for interpreting nanoscale friction, also in other materials. We emphasize that our interpretation of the frictional dissipation mechanism is distinct from other studies where viscous electronic damping of the tip motion has been discussed but also recognized as much too small to account for the observed effects. Future studies may take advantage of tailoring the phonon degrees of freedom in materials to better understand damping of slip pulses as well as to develop tactics to control friction.

## 4 Experimental Section

### 4.1 LSMO films.

The $x = 0.3$ film (La$_{0.7}$Sr$_{0.3}$MnO$_3$) was deposited on a (001) SrTiO$_3$ substrate with a La$_{0.4}$Sr$_{0.6}$MnO$_3$ buffer layer using the metal-aerosol deposition technique.[83] *Θ-2Θ* x-ray diffraction experiments (XRD) show no indications of any impurity phase and confirm highly oriented growth on the substrate (**Figure S1**). Small-angle x-ray scattering gives thicknesses of 5.8(2) nm and 5.4(2) nm for the $x =0.3$ film and buffer layer, respectively (**Figure S2**). The $x =0.3$ film undergoes the typical second-order phase transition from a ferromagnetic (FM) metal to paramagnetic (PM) metal which is accompanied by a change in the sheet film resistance with increasing temperature, which is characteristic of the formation of Jahn-Teller polarons[45-49]. The transition was characterized using SQUID magnetometry (**Figure S3**) and four-point sheet resistance measurements (**Figure S4**). The metal-metal transition temperature extracted from the resistivity data is $T_{MM} = max(\rho^{-1}d\rho/dT)$ = 330 K and the Curie temperature extracted from the magnetometry data is $T_C = max(d\mu/dT)$ = 338 K. According to the phase diagram,[84] this transition occurs in the

5.8 nm thick La$_{0.7}$Sr$_{0.3}$MnO$_3$ layer, while the buried La$_{0.4}$Sr$_{0.6}$MnO$_3$ buffer layer remains a PM-metal above room temperature and at all temperatures probed in this study. The specimen has an RMS-roughness of approximately 0.7 nm (**Figure S5**), extracted from 500 x 500 nm$^2$ AFM topography scans, which did not change as the temperature was cycled.

The $x$ =0.2 film (La$_{0.8}$Sr$_{0.2}$MnO$_3$) was deposited on a (001) SrTiO$_3$ substrate by sputter deposition.[51] $\Theta$-2$\Theta$ x-ray diffraction experiments show no indications of any impurity phases and confirm highly oriented growth on the substrate (**Figure S6**). The film thickness of 70 nm was determined by small-angle x-ray scattering (**Figure S7**). The film undergoes a phase transition from a ferromagnetic (FM) metal to paramagnetic (PM) metal which was characterized using SQUID magnetometry giving a Curie temperature at $T_C$ = $max(d\mu/dT)$ = 222 K (**Figure S8**). The resistance measurement associated with the phase transition was characterized using a temperature-dependent four-point measurement (**Figure S9**). This yielded a transition from the metal to polaronic state at $T_{MM}$ = $max(\rho^{-1}d\rho/dT)$ = 186.8 K. By tuning the deposition temperature and optimizing deposition parameters, the transition temperature was reduced by more than 100 K compared to the phase diagram,[84] presumably due to stresses from the epitaxial relation with the substrate and from sputter preparation-induced point defects. [51] The film shows an RMS-roughness of less than 0.2 nm, extracted from 500 x 500 nm$^2$ AFM topography scans (**Figure S10**).

**4.2 AFM measurements.**

AFM experiments were performed with a commercial Omicron VT-AFM/STM in a vacuum chamber at a base pressure of $p$ = 10$^{-10}$ mbar. The La$_{0.7}$Sr$_{0.3}$MnO$_3$ film ($x$ = 0.3) was radiatively heated from the back and the sample surface temperature was directly calibrated before experiments were performed. For measurements on La$_{0.8}$Sr$_{0.2}$MnO$_3$ ($x$ = 0.2), the sample was clamped to cryogenic stage whose temperature was controlled by adjusting the heating power of a resistor integrated into the stage and the liquid nitrogen flow. The surface temperature of the specimen is expected to be around 20 K warmer than the stage due to the poor thermal contact between the stage and the lightly clamped specimen.

Commercially available rectangular, single crystalline silicon cantilevers (Nanosensors PPP CONTSCR) with a nominal tip radius of less than 10 nm were used (although an scanning electron microscopy (SEM) image, **Figure S11**, suggests an actual radius of 21 nm). The normal $k_n$ and torsional $k_t$ cantilever spring constants used for the $x = 0.3$ measurements were obtained from the manufacturer values for the median cantilever dimensions and literature values for elastic constants ($k_n$ = 0.52 N m$^{-1}$ and $k_t$ = 42.74 N m$^{-1}$), while the normal and lateral forces were calibrated using the procedures described in references.[86-88] The La$_{0.8}$Sr$_{0.2}$MnO$_3$ specimen ($x$ = 0.2) was measured with an un-calibrated, pre-mounted cantilever of the same type as for the $x$ = 0.3 film. Estimates of the spring constants were obtained by assuming the average pull-off forces are the same for the $x$ = 0.2 and $x$ = 0.3 specimens (Figure 1(e)), leading to normal spring constants of $k_n$ = 0.03 N m$^{-1}$ and 0.02 N m$^{-1}$ for the first and second $x$ = 0.2 measurements, respectively. These values are lower than for the $x$ = 0.3 cantilevers presumably due to variations in cantilever dimensions, but are still within the manufacturer specifications. Assuming they result from a decreased cantilever thickness, we obtain values for the torsional spring constants for the cantilevers used in the first and second measurements of the $x$ = 0.2 film of $k_t$ = 15.02 N m$^{-1}$ and $k_t$ = 9.87 N m$^{-1}$, respectively.

Nanoscale friction measurements were performed using AFM-based lateral force microscopy. The lateral forces during sliding at a constant velocity and under a constant applied normal force are obtained from the measured torsion of the cantilever. By measuring the torsion during friction loops (trace and retrace scanning along the same line on the film surface, Figure 1(a)), the effects of topography are largely separated out from friction effects.[54] The friction forces $F_f$ were obtained by averaging 100 or 256 friction loops that were performed in 100 x 100 nm$^2$ or 500 x 500 nm$^2$ regions on the surface of the specimens for each normal load and temperature of interest. To minimize wear during the friction experiments, the applied normal forces $F_N$ were kept below 30 nN and any possible changes in the contact were monitored through tip-sample adhesion measurements. Adhesion forces were obtained by recording 100 force-distance curves before and after probing the frictional properties at each temperature and determined by averaging the extracted pull-off forces.

**Supporting Information**

Supporting Information is available from the Wiley Online Library or from the author.

**Acknowledgements**

This work is funded by the Deutsche Forschungsgemeinschaft (DFG, German Research Foundation) 217133147/SFB 1073, project A01. The authors thank P. Ksoll and D. Schwarzbach for providing four-point measurements to characterize the phase transition in $La_{0.8}Sr_{0.2}MnO_3$. Furthermore, we thank O. Gordon for providing a Python interface to perform partially automated measurements. Our thanks also go to R. L. Vink, S. Merten, T. Brede and F. Schönewald for discussions of the results and to Alec Wodtke for allowing us to use the vacuum AFM system.

# Supporting Information

# Polaronic Contributions to Friction in a Manganite Thin Film

Niklas A. Weber[1], Dr. Hendrik Schmidt[1], Tim Sievert[1], Prof. Christian Jooß[1], Dr. Friedrich Güthoff[2], Prof. Vasily Moshneaga[3], Prof. Konrad Samwer[3], Prof. Matthias Krüger[4], and Prof. Cynthia A. Volkert[1,*]

[1] Institute of Materials Physics, University of Göttingen, 37077 Göttingen, Germany

[2] Institute of Physical Chemistry, University of Göttingen, 37077 Göttingen, Germany

[3] 1. Physics Institute, University of Göttingen, 37077 Göttingen, Germany

[4] Institute for Theoretical Physics, University of Göttingen, 37077 Göttingen, Germany

E-mail: volkert@ump.gwdg.de

**Section S1. Sample Characterization**

θ-2θ x-ray diffraction experiments (XRD) and x-ray reflectometry (XRR) were carried out on the $La_{0.7}Sr_{0.3}MnO_3$ and $La_{0.8}Sr_{0.2}MnO_3$ specimens using a Bruker D8 with Cu-$K_\alpha$ source (**Figure S1, S2, S6, and S7**). The XRD spectra show (001) family LSMO reflections superimposed on strong reflections from the (001) oriented single crystal $SrTiO_3$ substrate (Figure S1 and S6), confirming the heteroepitaxial relation between the film and substrate. Diffration peaks from the aluminum sample holder peaks can also be seen. The XRR spectra (Figure S2 and S7) were fit to obtain the thicknesses of the deposited LSMO films.[S1]

The magnetic and electrical properties of the specimens were measured using SQUID (superconducting quantum interference device) and four-point resistivity (**Figure S3, S4, S8, and S9**) and the transition temperatures determined using the relations $T_C = max(d\mu/dT)$ and $T_{MM} = max(\rho^{-1}d\rho/dT)$.

Topography maps (**Figure S5 and S10**) were obtained using standard contact AFM methods with a commercial Omicron VT-AFM/STM in a vacuum chamber at a base pressure of $p = 10^{-10}$ mbar.

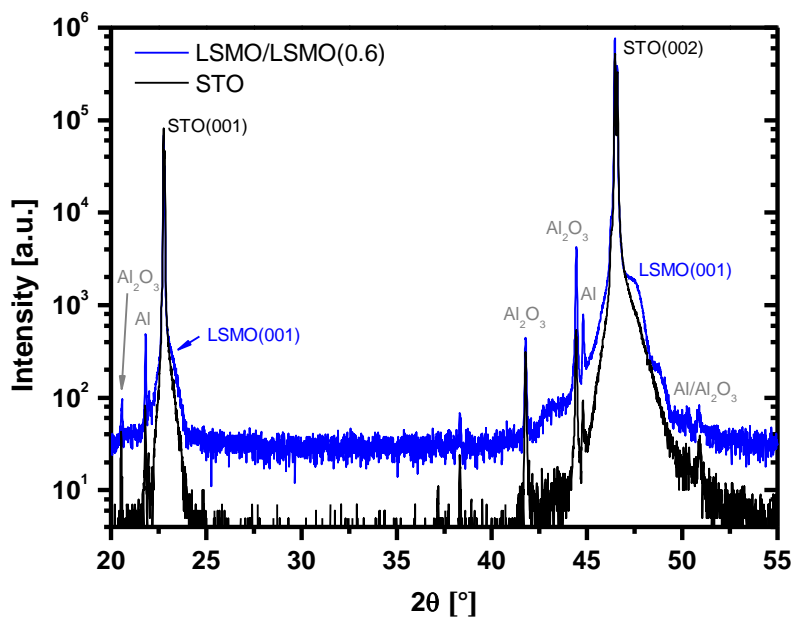

**Figure S1.** θ-2θ XRD measurements of the La$_{0.7}$Sr$_{0.3}$MnO$_3$/La$_{0.4}$Sr$_{0.6}$MnO$_3$ bilayer which was deposited on a (001) oriented SrTiO$_3$ substrate. The lattice constants were determined to be 3.871(3) Å und 3.781(2) Å, respectively.[S2]

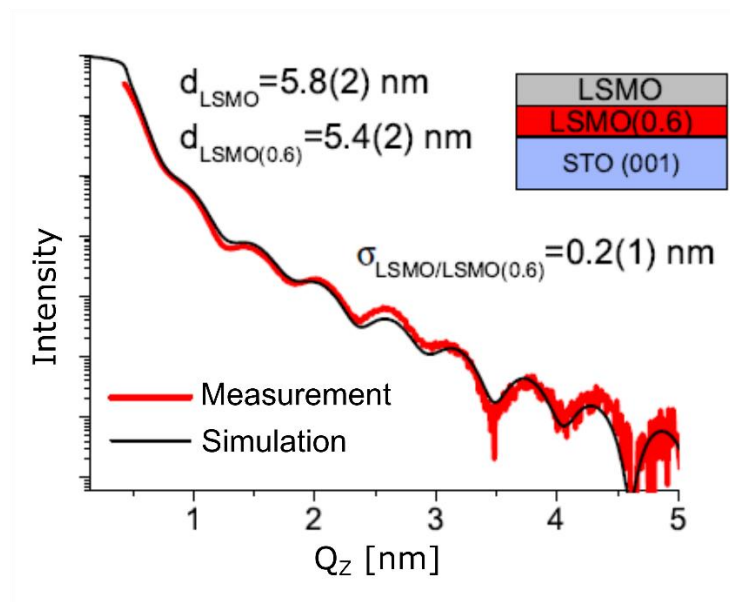

**Figure S2.** XRR measurements of the La$_{0.7}$Sr$_{0.3}$MnO$_3$/La$_{0.4}$Sr$_{0.6}$MnO$_3$ bilayer film yield thicknesses of 5.8 and 5.4 nm, respectively.[S2]

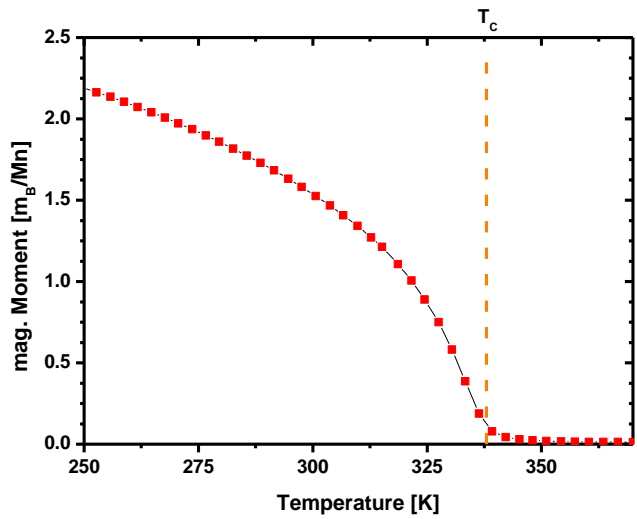

**Figure S3.** SQUID magnetometry measurements carried out on $La_{0.7}Sr_{0.3}MnO_3/La_{0.4}Sr_{0.6}MnO_3$ yield a Curie Temperature of $T_C$ = 338 K.

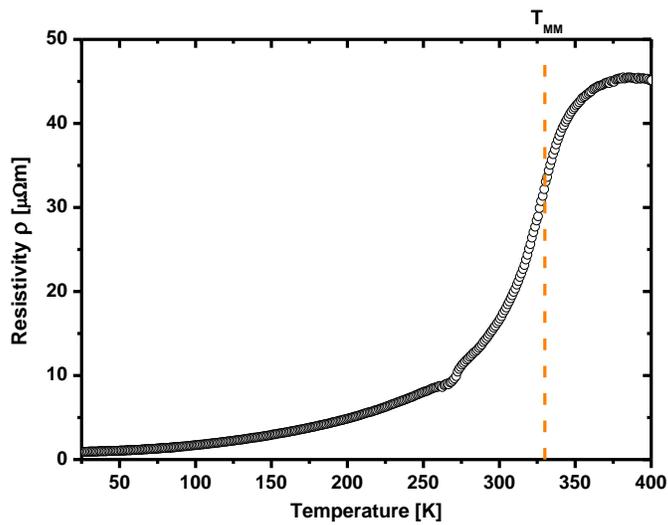

**Figure S4.** $La_{0.7}Sr_{0.3}MnO_3/La_{0.4}Sr_{0.6}MnO_3$ film resistance measured in a four-point geometry. The metal-metal transition temperature is found at $T_{MM}$ = 330 K.

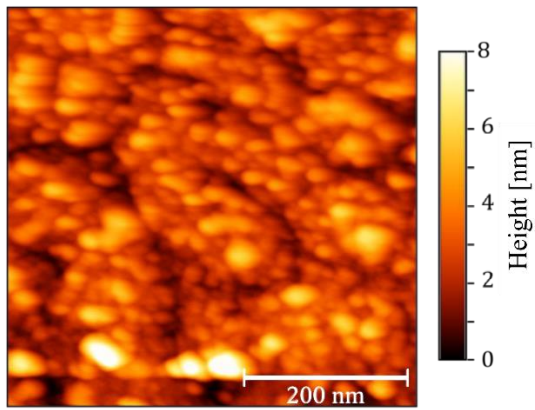

**Figure S5.** Topography map of $La_{0.7}Sr_{0.3}MnO_3$ measured at room temperature under UHV conditions using an Omicron VT-AFM/STM. The specimen shows an RMS-roughness of approximately 0.7 nm. No effect of temperature on surface morphology was observed.

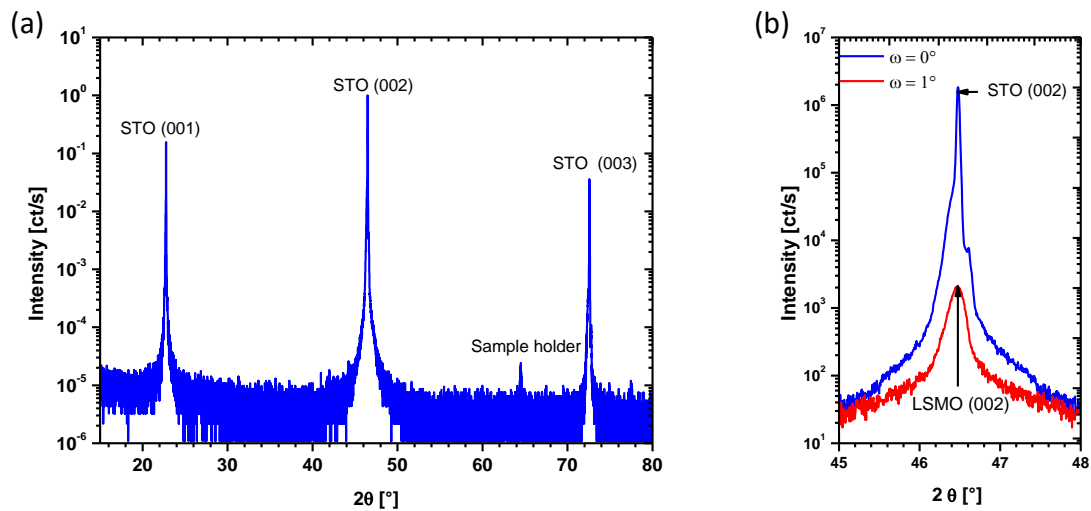

**Figure S6.** (a) θ-2θ XRD measurements of $La_{0.8}Sr_{0.2}MnO_3$ deposited on a (001) oriented $SrTiO_3$ substrate. The peaks from the $La_{0.8}Sr_{0.2}MnO_3$ are superimposed on the STO (001) peaks. (b) To distinguish substrate from film a second measurement for 2θ between 45° and 50° at ω=0° and 1° were conducted. A lattice constant of 3.906(2) Å was determind from the spectra.

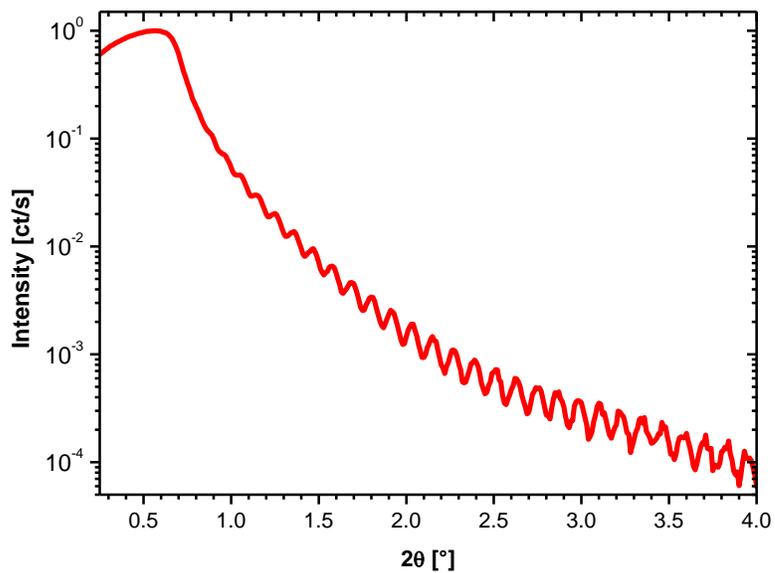

**Figure S7.** XRR measurements carried out on sputtered La$_{0.8}$Sr$_{0.2}$MnO$_3$ on SrTiO$_3$ yield a film thickness of 70 nm.

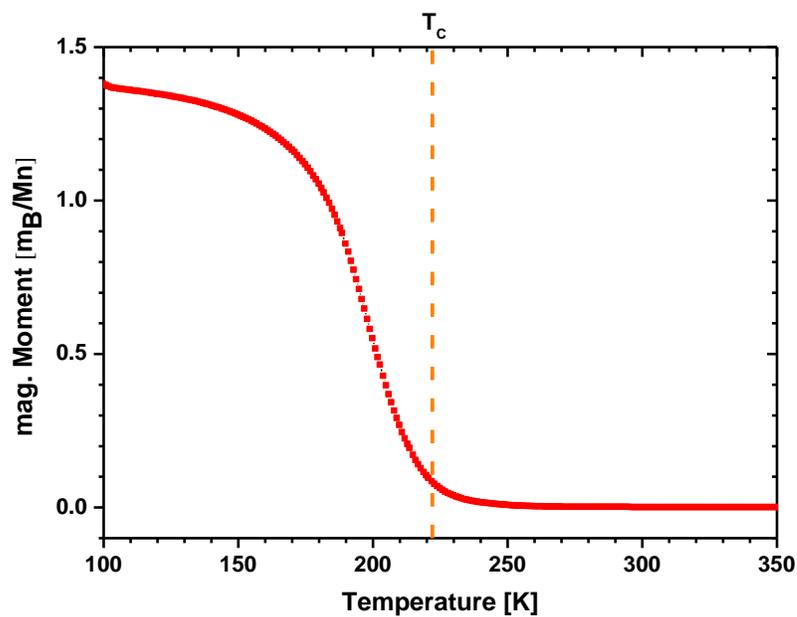

**Figure S8.** SQUID magnetometry measurements carried out on La$_{0.8}$Sr$_{0.2}$MnO$_3$ yield a Curie Temperature of T$_C$ = 222 K.

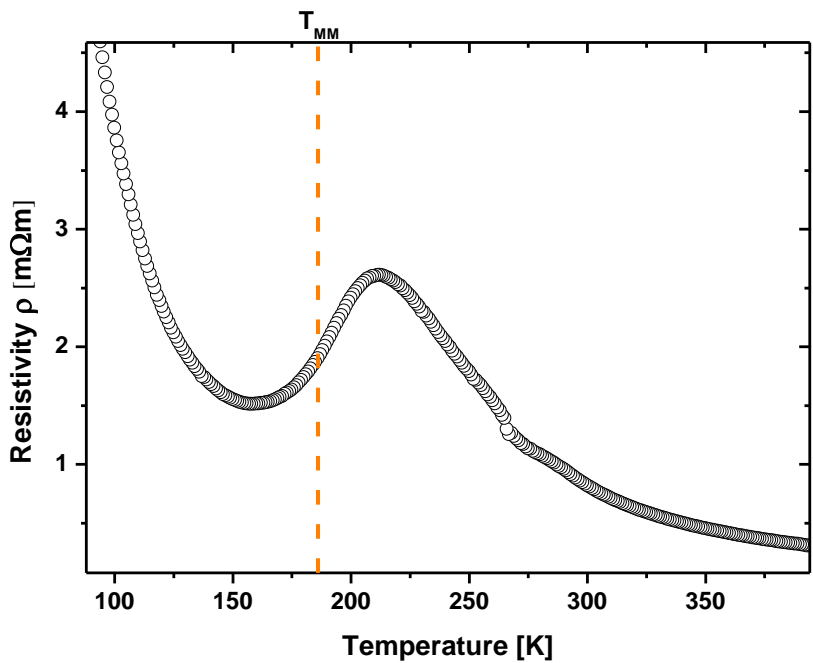

**Figure S9.** Four-point resistivity measurements on La$_{0.8}$Sr$_{0.2}$MnO$_3$ reveal a metal-metal transition temperature at T$_{MM}$ = 187 K.

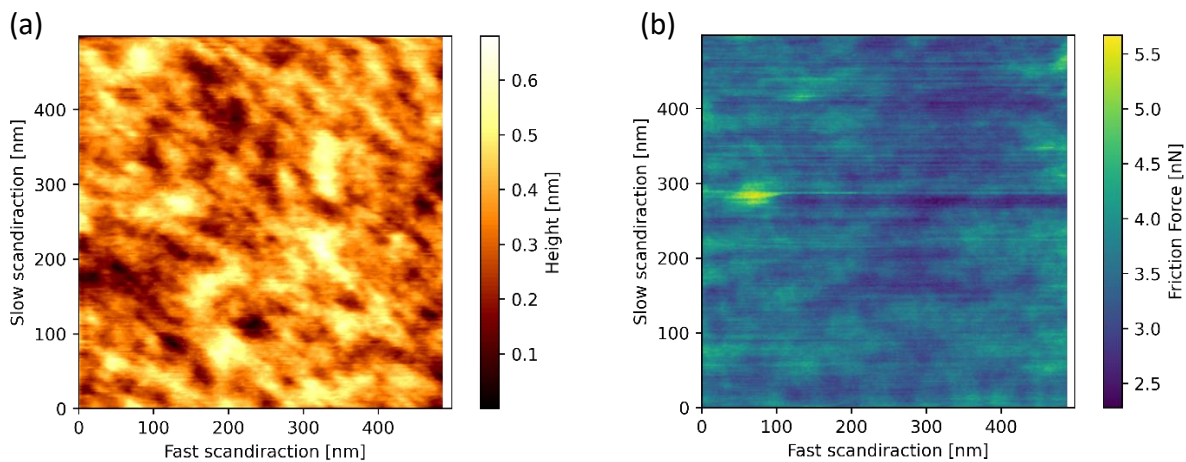

**Figure S10.** Topography (a) and corresponding friction map (b) calculated from lateral trace and retrace data from La$_{0.8}$Sr$_{0.2}$MnO$_3$ measured at room temperature at a constant normal force of $F_N = 0.7$ nN. The RMS roughness remained constant below 0.2 nm during the entire temperature series friction measurements. No effect of temperature on surface morphology was observed.

**Section S2. Cantilever Characterization**

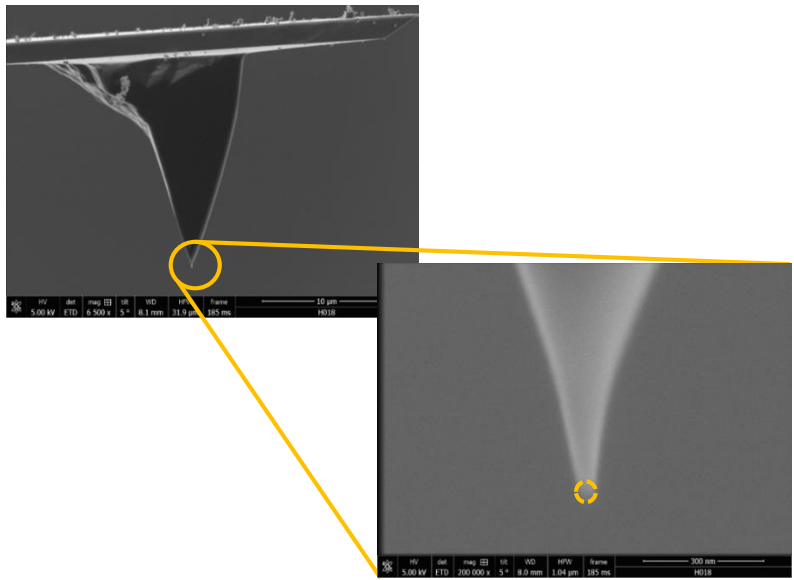

**Figure S11.** Scanning electron microscopy images from the tip of a cantilever (Nanosensors PPP-CONTSC) allowed an estimation of the tip radius as 21 nm.

**Section S3. Electrical Characterization using C-AFM**

Conductive-AFM (C-AFM) measurements were carried out under UHV conditions using Pt-coated silicon cantilevers (Nanosensors PPP-CONTSCPt) with similar spring constants to the Si cantilever used in the friction study to gain a deeper understanding of the nanoscale electrical properties of the $La_{0.7}Sr_{0.3}MnO_3/La_{0.4}Sr_{0.6}MnO_3$ bilayer. In a first step, we recorded several I-V curves in the ferromagnetic state at room temperature and in the paramagnetic state at $T$= 380 K. As shown in *Figure S*the sample becomes more insulating in the high temperature phase. We attribute the insulating behavior at low currents to the presence of a highly insulating, electrically dead surface layer, which is often reported for perovskite manganites.[S3]

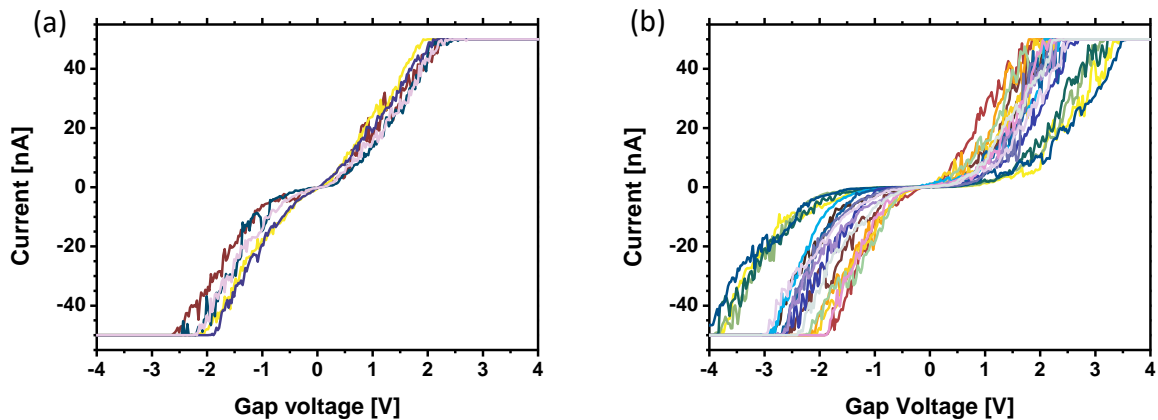

Figure S12. I-V spectroscopy curves recorded at (a) room temperature where the sample is a ferromagnetic metal and (b) at T= 380 K where it is a paramagnetic metal. There is a clear increase in the resistance of the high temperature phase.